\documentclass[final]{svjour2}
\usepackage{amsmath}
\usepackage{amsfonts}
\usepackage{graphicx}
\usepackage{rotating}
\usepackage{amssymb}
\usepackage{mathptmx}
\usepackage[numbers]{natbib}
\makeatletter
\journalname{Journal of Low Temperature Physics}

\bibpunct{}{}{,}{s}{}{,}

\begin{document}

\newcommand{\hdblarrow}{H\makebox[0.9ex][l]{$\downdownarrows$}-}
\title{In Pursuit of the Elusive Supersolid}

\author{X. Mi$^1$ \and J.D. Reppy$^1$}

\institute{1:Laboratory of Atomic and Solid State Physics\\Cornell University, Ithaca, NY, USA 14850-2501\\
\email{jdr13@cornell.edu}}

\date{07.10.2013}

\maketitle


\begin{abstract}

The excitement following the initial report of supersolid behavior for $^4$He embedded in porous Vycor glass has been tempered by the realization that many of the early supersolid observations were contaminated by effects arising from an anomaly in the elastic properties of solid $^4$He. In an attempt to separate dynamic elastic effects from a true supersolid signal, we employed a torsional oscillator with two eigen-frequencies to study the $^4$He-Vycor system. We found that frequency-dependent elastic signals can entirely account for the observed period shift signals. Although, we conclude that supersolid does not exist for the $^4$He-Vycor case, the question of its presence in bulk samples remains open. In our current experiments we  apply the two-frequency test to bulk samples of solid $^4$He. Again we find a frequency-dependent contribution arising from elastic effects; however, in some cases we also find a small frequency-independent contribution, which may indicate the existence of a remnant supersolid phase. Given the history of this subject such results must be treated with caution.

\end{abstract}

\section{Introduction}
In 2004 Kim and Chan (KC) reported \cite{B1,B2} an anomalous drop in the period of a torsional oscillator (TO) containing solid $^4$He at temperatures below 200 mK.  The anomalous period signal was interpreted as evidence for a supersolid state in the solid and was quickly replicated in a number of other laboratories \cite{B3,B4,B5,B6}.  Interpreted as a supersolid phenomenon, the anomalous TO period shift is seen as a consequence of a superfluid-like decoupling of a fraction of the solid moment of inertia from the TO. As in the case of the superfluid, in a supersolid scenario, a supersolid fraction can be defined as the ratio of the anomalous period shift to the total period shift ascribed to the addition of the moment of inertia of the solid. Unfortunately, in the decade since the first observation of the anomalous TO period signals, the initial excitement engendered by this discovery has been tempered by the gradual accumulation of experimental evidence that suggests an alternative explanation for the observed TO signals.

In this paper we shall discuss the development of experimental evidence calling into question a supersolid interpretation of the anomalous TO period signals. An important element in this story is the study \cite{B7} in 2007 by Day and Beamish (DB) of the low temperature elastic properties of solid $^4$He. DB observed an anomalous increase in the shear modulus of the solid over the same temperature range as the TO supersolid phenomenon. In addition, they observed a sensitivity to $^3$He concentration similar to that found by KC in their early TO experiments.

\section{Interplay between Solid $^4$He Elasticity and TO Dynamics}

Two approaches have been pursued in the attempt to calculate the magnitudes of the shifts in the TO period due to influence of changes in the shear modulus. One, most actively pursued by Chan's Penn State group, has been a finite element analysis \cite{B8} of the frequency dependent elastic response of the TO. A second approach has been through an analytic analysis \cite{B9,B10,B11} of various simplified geometries for the TO. This second approach has the advantage that one can understand the relative importance of various aspects of the TO design in determining the magnitude of the TO sensitivity to changes in the elastic properties of the solid $^4$He sample. Both approaches have shown that elastic effects can seriously contaminate the period shift signals that have been observed for a number of TO experiments.

There are two principal ways in which changes in the $^4$He shear modulus lead to shifts in the TO period. In most TO’s it has been the practice to fill the cell through a fill line drilled up through the center of the cell torsion rod. The solid $^4$He in the torsion rod then contributes to the total torsion constant for the cell, and thus an increase in the shear modulus of the solid will lead to a decrease in the TO period. This effect was first exploited by Paalanen, Bishop and Dail \cite{B12} in their 1981 study of the low temperature elastic anomaly of solid $^4$He. More recently, Beamish {\it et al.} \cite{B13}, have revisited this problem. They find that the fractional shift in the TO period, $\Delta P/ P$, is given by $\Delta P/ P\approx - \frac{1}{2} (\Delta \mu_{\text{He}} / \mu_{\text{Rod}})(r_{\text{i}}^4/r_{\text{o}}^4)$, where $\Delta \mu_{\text He}$ is the change in shear modulus of the solid in the torsion rod and $r_{\text i}$ and $r_{\text o}$ are the inner and outer radii of the torsion rod. It should be noted that the right hand side of this equation does not depend on the TO period. A similar situation holds for the fractional period shift, $\Delta P / P = (1/2)(\Delta I/ I_{\text{He}})$, which might result from a small change, $\Delta I$, in the total cell moment of inertia such as that due to a possible supersolid decoupling.

The fractional period or frequency sensitivity of the TO to the torsion rod effect depends on the ratio of the inner to the outer radii of the torsion rod raised to the fourth power. Thus, by reducing this ratio, the influence of the solid in the torsion rod on the TO period can be made insignificant as compared to observed period shifts. Unfortunately there are a number of experiments where this ratio is not small and the observed period shifts reflect, in large part, changes in the shear modulus of the solid $^4$He. Beamish {\it et al.}\cite{B13} have summarized the degree of influence of the torsion rod effect for a number of different TO experiments.

Another route by which changes in the shear modulus affect the TO period occurs when the solid plays a dynamic role in coupling individual mass elements of the TO to the external torque applied by the torsion rod. An example would be the case where the mass element is the inertia of the solid sample itself.  Here, the force required for the acceleration of the solid during the oscillatory motion arises from the strain field in the solid associated with the differential motion between the solid and the cell walls. The relative average motion of the solid and the walls will be proportional to the acceleration and thus vary as the square of the frequency and will also be inversely proportional to the shear modulus \cite{B10}. As a result of this relative motion, the solid, on average, rotates though a larger total angle than the more rigid walls of the cell and acts as an effective increase in the moment of inertia of the TO, with an increased period compared to the completely rigid case. Here, as in the case of solid in the torsion rod, an increase in the shear modulus will lead to a decrease in the TO period. There is, however, an important difference; in this dynamic case, the magnitude of the fractional period shift will be proportional to the square of the frequency. Thus, a double frequency TO can be employed to differentiate between a true, frequency-independent, supersolid signal and a signal arising from dynamic effects due to temperature variations in the elastic modulus of the solid. A more extensive discussion of the frequency-squared effect for a number of cell geometries is given in Ref. 10.

In addition to the model calculations, there has been an accumulation over time of experimental evidence that calls for a degree of caution in interpreting the observed TO period shifts as conclusive evidence for a supersolid state. An early warning sign came in 2006 with the experiments of Rittner and Reppy \cite{B3} (RR). In these experiments a bulk $^4$He solid sample was contained in a cubic volume designed to eliminate the possibility of solid slippage at the cell boundary, as had been suggested in 2004 by Dash and Wettlaufer\cite{B14}. This experiment, in addition to confirming the original KC results, found that the apparent supersolid period shift signals were sensitive to annealing and could be progressively reduced in magnitude, even below the level of observation, by sufficient annealing. RR also found that a sizable signal could be restored by the melting of the sample followed by rapid refreezing. These results demonstrated that that the apparent supersolid signals were not a universal property of solid $^4$He, but depend in some way on the degree of disorder present in the sample. At Cornell, the Reppy group continued to pursue a line of experiments designed to follow up on the idea that disorder played an important role in the supersolid phenomenon. This line of research culminated in a 2010 experiment \cite{B15} that allowed an {\it in situ} increase in the degree of disorder in the solid sample through plastic deformation of the solid. Following plastic deformation of the sample, there was a marked increase in the overall magnitude of the period shift signal, as one might have expected on the basis of an increase in the degree of disorder. The surprising result, however, was the observation that the change in the TO period occurred in the high temperature region and that at the lowest temperature there was essentially no change in the TO period following the deformation of the sample. This is quite the opposite of the behavior one would expect for a superfluid-like signal, where the change in TO period would be largest at the lowest temperature and the high temperature data above the transition temperature would remain unchanged. The conclusion implied by this experiment is that the observed signal must have a nonsuperfluid origin. For a discussion of the implications of this experiment see the commentary by John Beamish \cite{B16}. 

\section{Multiple Frequency TO's}

\begin{figure}
\begin{center}
\includegraphics[%
  width=0.8\linewidth,
  keepaspectratio]{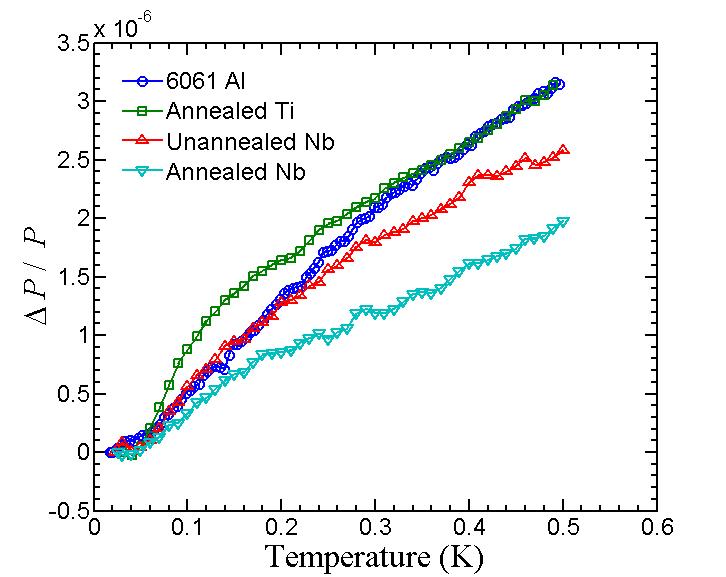}
\end{center}
\caption{(Color online) Empty cell period shift with respect to 20 mK value, $\Delta P$, divided by the total period $P$, of TOs made with different torsion rod materials. The Ti torsion rod has been heat treated in air at 600 $^\circ$C for 2 hours post-machining. The Nb torsion rod was annealed at 1000 $^\circ$C for 2 hours in vacuum.}
\label{fig1}
\end{figure}

Motivated by this unexpected result, we developed a series of multiple frequency TO’s that would allow a discrimination between a true supersolid signal and the period shifts arising from the temperature-dependent elastic anomaly. We have undertaken a program of progressive improvement in the performance of our TO's. To this end we have examined the temperature-dependence for the TO period that results from the temperature-dependent shear modulus of the torsion rod itself. In Fig.~\ref{fig1} we show the fraction period shift background as a function of temperature for a number of commonly used torsion rod materials including annealed Ti, Nb annealed and unannealed, as well as 6061 Al. It was found that annealed Nb has the lowest dissipation and also smallest temperature dependence in its shear modulus. However, we have found that Nb is prone to large irreversible changes in its mechanical properties when subjected to thermal cycling and mechanical stress, which makes it an unreliable candidate for the torsion rod. Ti has the disadvantage of having an abrupt change in the temperature dependence of its shear modulus near 300 mK, possibly associated with the transition into a superconducting state. Therefore, we have chosen 6061 Al as the material for our TOs because of its resistance to the effects of thermal cycling and mechanical stress, as well as the relatively linear temperature dependence of its shear modulus.

For our latest multiple frequency oscillators we have reduced the diameter of our fill lines running up through the center of the torsion rods to 0.01 cm to give a ratio to our typical outer diameter of about 50. In addition we have made the base of the TO structure more massive to reduce a subtle effect, noted by Maris \cite{B11}, where elastic properties of the solid in contact with the TO baseplate contribute to the effective torsion constant of the torsion rod. The “Maris” effect can be important in cell designs where the bottom plate connected to the torsion rod is designed to be very thin in the interest of raising the sensitivity of the TO to mass-loading. Such designs therefore have the problem of inducing large signals due to changes in solid $^4$He shear modulus. It is worth noting that the signal arising from the inertial acceleration of the solid sample can also be minimized by confining the sample between the walls of an annular geometry or in the interconnected pores of a rigid porous medium such as Vycor glass.

In one application of the multiple frequency technique \cite{B18}, we repeated KC’s original supersolid experiment \cite{B1}, where the solid $^4$He sample was confined in porous Vycor glass. Our results were very similar to KC’s. However, in our case we had two different frequencies available and expected to be able to distinguish a dynamic signal and a frequency-independent contribution arising from a true superfluid or supersolid. As it turned out, the entire signal could be ascribed, within error, to a dynamic effect with the characteristic frequency-squared dependence.  Our conclusion was then that there was no sign of supersolid in the Vycor sample. At the time of the publication of this result, we did not understand the origin of the dynamic signal; Moses Chan, however, emphasized to us that the origin probably arose from the thin layer of bulk solid at the bottom of the cylindrical Vycor sample. Indeed, finite element calculations at Penn State and our own model calculations confirm this to be the correct explanation. Simply stated, the solid in this layer provides part of the oscillatory force required for the acceleration of the upper part of the cell including the Vycor sample and thus leads to a signal proportional to the square of the frequency. In a beautiful experiment \cite{B19}, Chan's Penn State group has examined the TO response for solid $^4$He in Vycor in a cell where bulk solid was almost entirely excluded and found a complete absence of any supersolid signal.

The collapse of the case for supersolid in Vycor suggests that supersolid may not exist. There are, however, experiments such as the rotation experiments \cite{B20,B21}, where the period shift signal is reduced as the rotation speed is increased, for which no one has as yet provided an alternative explanation. It would seem important, in view of the absence of understanding, to apply the two-frequency test to these putative supersolid signals.

\begin{figure}
\begin{center}
\includegraphics[%
  width=0.8\linewidth,
  keepaspectratio]{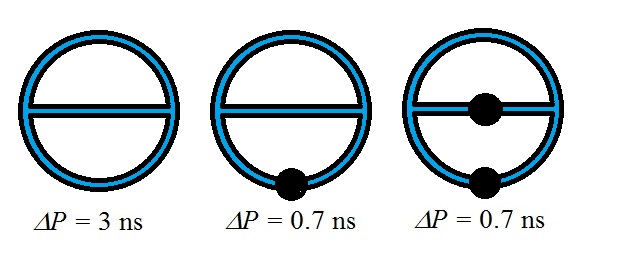}
\end{center}
\caption{(Color online) The blocked annulus experiment: Blue channels indicate the annular region of the cell filled with solid $^4$He. Black circles represent 1.5 mm copper rods sealed with epoxy, blocking the open channel. The width of the channel is 0.5 mm.}
\label{fig2}
\end{figure}

An appreciation for the role played by elastic effects can provide an alternate explanation for the ``blocked'' annulus experiments that were originally considered as strong supporting evidence for long-range superflow in the solid. In the original experiment of this type, performed by KC \cite{B2} and later repeated at Cornell \cite{B17}, a substantial period shift signal was observed for a TO where the sample was contained in a narrow annular channel. KC then constructed a second cell of similar dimensions where any possible superflow around the annulus was blocked by a partition. For the blocked cell only a small signal was observed, which was attributed to potential backflow of the superfluid.

Following up on these initial results a more sophisticated version of the blocked annulus experiment was attempted at Cornell with the help of Wonsuk Choi, a visiting graduate student from E. Kim's group at KAIST. The basic geometry for the cell consisted of a narrow open annular channel with a cross channel along a diameter connecting opposite sides of the annulus. When superfluid or supersolid is introduced into this geometry, as a first step, only the superfluid in the cross channel is entrained in the oscillatory motion of the cell. Next, a blocking partition is place at one point in the annular channel as shown schematically in Fig.~\ref{fig2}. With the introduction of this barrier a substantial fraction of the superfluid moment of inertia is coupled to the oscillator; however, a portion of the superfluid still remains free to flow around the noncircular D-shaped path formed by the top half of the circular annulus and the cross channel. In a third step, a blocking partition is placed at the center of the cell, blocking any superflow around the D-shaped flow channel and coupling almost the entire superfluid moment of inertia to the TO. These steps are shown in Fig.~\ref{fig2}, along with the observed period shifts at each step.

Before the first block was introduced, a sizable period shift signal of about 3 ns was observed for the solid. With the first block in place, the remaining superflow path was around the D-shaped region and the solid period shift signal was reduced to about 0.7 ns, as expected on the basis of potential flow considerations. The surprise came when, after blocking the D flow path at the center of the cross channel, the period shift signal was unchanged, remaining at the 0.7 ns level. At the time, 2009, when this experiment was performed, we were at a loss to explain this result which seemed in complete contradiction to our expectations based on a superfluid-like model for the supersolid state.

Today, in light of our increased appreciation for the role played by elastic modulus of the solid in TO behavior, both the initial blocked-annulus experiments of KC and RR and the 2009 D-shaped geometry experiment can best be understood in terms of an interplay between the elastic anomaly and the structure of the TO cell. In an open annular cell, before a blocking partition is introduced, the solid can play a significant role in coupling various parts such as the inner and outer walls of the annulus of the cell together, thus providing an observable period shift signal. The effect of a partition placed across the annular channel is not only to block possible flow around the annulus but also to provide a much more rigid clamping of the inner to the outer wall of the cell and thus substantially reduce the contribution of the elastic anomaly to the TO period. In the case of the D-shaped experiment, the block was placed at the center of the cell where the motion is essentially zero, so the contribution of the solid elasticity to the TO behavior is unchanged.

\begin{figure}
\begin{center}
\includegraphics[%
  width=0.8\linewidth,
  keepaspectratio]{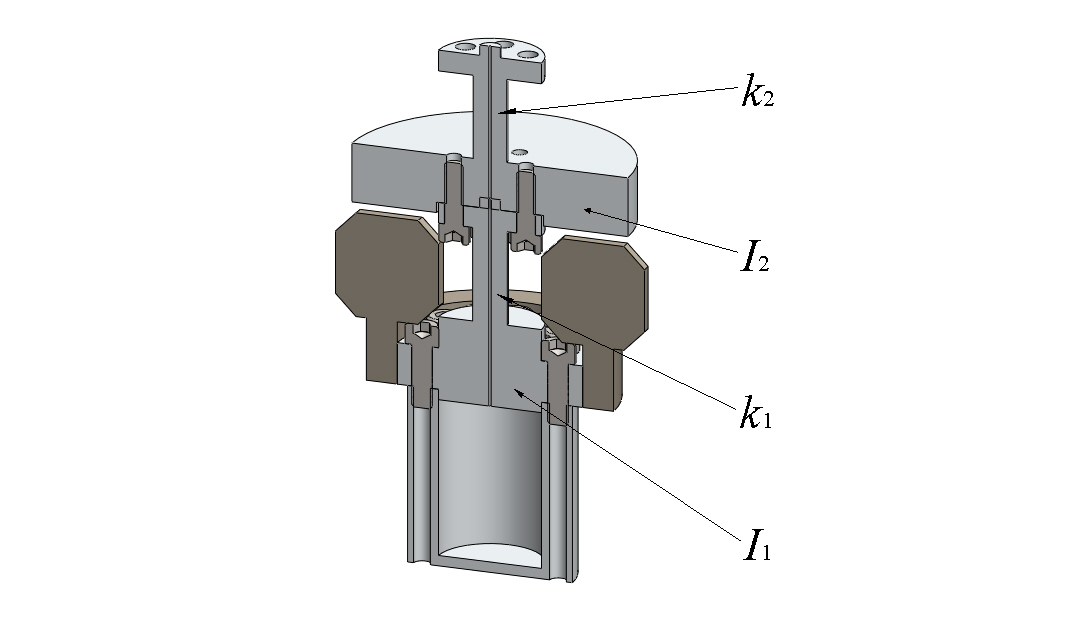}
\end{center}
\caption{The double frequency oscillator.}
\label{fig3}
\end{figure}

In our current work we are attempting to improve the quality of our double frequency measurements in order to make a clear identification of any frequency independent contribution to the period shift signal data that might signal the existence of a supersolid component. In one experiment, we have examined the period signals for a double oscillator with frequencies of 446 Hz and 1155 Hz. A picture of the cell is shown in Fig.~\ref{fig3}. The solid sample is contained in a cylindrical volume with a radius of 0.794 cm and a height of 2.381 cm. With these dimensions, we expect to be able to resolve the frequency dependent “inertial” signal arising from the acceleration of the solid $^4$He sample within the cylindrical sample volume. If we assume a 10\% shift in the solid $^4$He shear modulus, then following Ref. 10, Sec. 3.3, we estimate the period shift due to the solid $^4$He inertial term to be $\Delta P_- = 0.086$ ns and $\Delta P_+ = 0.088$ ns. The inner diameter of the fill lines through the torsion rods is 0.033 cm to give a ratio of 15 between the outer and inner diameter of the rods. For this ratio we would expect that a 100\% change in the shear modulus of the solid in the fill line would give a frequency independent change in period on the order of $\Delta P / P = 4.5 \times 10^{-8}$. The bottom of the sample cavity adjoining the torsion rod $k_1$ is made to be very thick in an attempt to minimize the modification of the cell torsion constant, $k_1$, through the distortion of the bottom plate as described in Ref. 11. The expected signal due to the distortion of the end plate of the cell, based on 100\% change in the shear modulus of solid $^4$He sample, is on the order of $\Delta P / P = 2 \times 10^{-7}$. The mass loading sensitivities for the two modes are obtained from the observed period shifts upon melting the sample at low temperature. For the low frequency mode the period shift on melting is, $\Delta P_- = 3.126 \times 10^{-6}$ s, and for the high mode the shift is $\Delta P_+ = 0.477 \times 10^{-6}$ s. 

\begin{figure}
\begin{center}
\includegraphics[%
  width=0.8\linewidth,
  keepaspectratio]{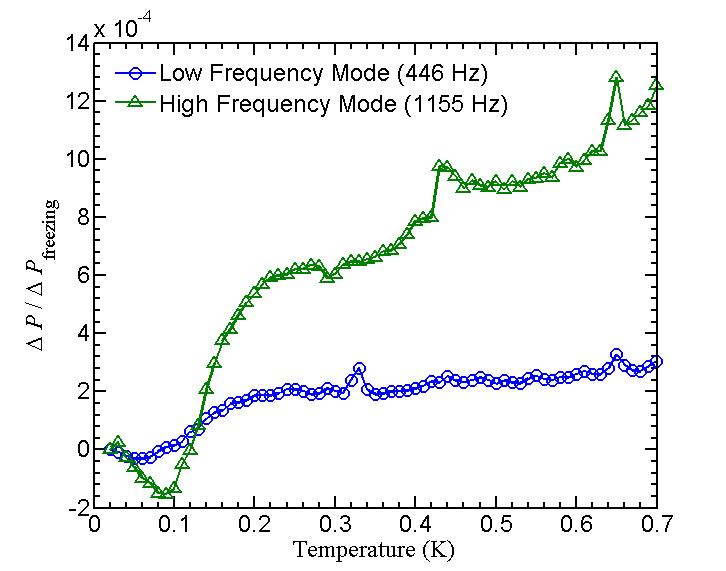}
\end{center}
\caption{(Color online) The normalized fraction, $\Delta P / \Delta P_{\text{freezing}}$, for the two modes, is plotted against temperature for the two modes. The quantity $\Delta P$ is the period shift of the filled cell with respect to 20 mK value and $\Delta P_{\text{freezing}}$ is the period shift of the cell seen upon freezing or melting of the solid $^4$He sample.}
\label{fig4}
\end{figure}

We now proceed to analyze the data from our two-frequency cylindrical cell along the same lines as in the Vycor case. Our basic data in this experiment consist of the period shifts for the two modes, obtained with the cell filled with a solid sample, $P_\pm(T)$,  and background data,  $P_{{\text B}\pm}(T)$, obtained with the cell empty. Subtracting the empty cell data from the full cell data, we have $\Delta P_\pm(T) = P_\pm(T) - P_{{\text B}\pm}(T)$, where the empty cell background periods, $P_{{\text B}\pm}(T)$, have been individually shifted by constant amounts to make $\Delta P_\pm(T = 0) = 0$. Normalized or fractional period shifts for the modes are defined as, $\delta P_\pm(T) = \Delta P_\pm(T) / \Delta P_{{\text{freezing}}\pm}$, where the period shifts, $\Delta P_{{\text{freezing}}\pm}$, are the values for the two modes observed at the melting or freezing of the sample. In our analysis we shall assume that the shear modulus does not itself depend on the frequency over the frequency range between the two TO modes. Day and Beamish \cite{B7} report a 1.5 to about 2\% for a variation in the shear modulus for a factor of 10 in frequency between 20 and 200 Hz. The ratio between the mode frequencies in this experiment is only a factor of 2.6, so we shall, therefore,  neglect this small dependence of the shear modulus on frequency in our analysis. In Fig.~\ref{fig4} we show the fractional data for each mode for a temperature range up to 0.7 K. If these data corresponded to a traditional superfluid signal then we would expect the fractional data for the two modes to lie on top of each other. Clearly they do not.

\begin{figure}
\begin{center}
\includegraphics[%
  width=0.8\linewidth,
  keepaspectratio]{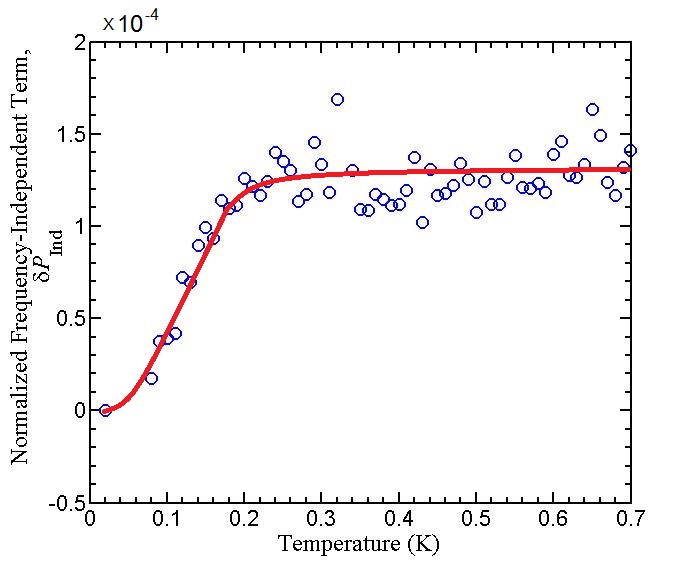}
\end{center}
\caption{(Color online) The normalized frequency-independent term, $\delta P_{\text{Ind}}$, calculated from the data of Fig.~\ref{fig4}, is plotted against temperature.}
\label{fig5}
\end{figure}

In the further analysis of the data, we also assume a model where the normalized period shifts can be expressed as the sum of two terms, one a frequency-independent part, $\delta P_{\text{Ind}}(T)$, and contributions from frequency-dependent terms, $\delta P_{\text{Freq}-} = a(T)f_-^2$ and $\delta P_{\text{Freq}+} = a(T)f_+^2$. The temperature-dependent quantity, $a(T)$, will depend on the temperature dependent shear modulus and the design of the TO cell. We then have two equations: $\delta P_-(T, f) = \delta P_{\text{Ind}}(T) + a(T)f_-^2$ and $\delta P_+(T, f) = \delta P_{\text{Ind}}(T) + a(T)f_+^2$. Solving for the frequency-independent term, we obtain $\delta P_{\text{Ind}}(T) = [r_- \delta P_- - r_+ \delta P_+]$, where we have introduced the notation, $r_- = [ P_-^2/ (P_-^2 - P_+^2)]$ and $r_+ = [ P_+^2/ (P_-^2 - P_+^2)]$. In Fig~\ref{fig5}, we plot this frequency-independent contribution as a function of temperature. The curve through the data is a polynomial fit. In the supersolid scenario these data would indicate a maximum supersolid fraction of $1.6 \times 10^{-4}$ at the lowest temperature. Although it would be gratifying to take these data as evidence of the supersolid state, the history of this subject dictates that caution is advised.

Our current efforts are directed at improving the quality of the double oscillator measurements through a search for better torsion rod materials and enhanced vibration isolation.

\begin{acknowledgements}
We would like to thank A.S.C. Ritter and W. Choi for assistance with the blocked annulus experiments and E.J. Mueller for theoretical advice.  This work has profited from the advice and criticisms of M.H.W. Chan, J. Beamish, and J.V. Reppy and is supported by the National Science Foundation through Grant DMR-060586, DMR-0965698 and CCMR Grant DMR-050404.
\end{acknowledgements}

\pagebreak

\end{document}